\documentclass[pdflatex,sn-aps,sn-mathphys,iicol]{sn-jnl}

\usepackage{amsmath}
\usepackage{graphicx}
\usepackage{epsf}
\usepackage{psfrag}
\usepackage{epsfig}
\usepackage{graphics}
\usepackage{amsfonts}
\usepackage{slashed}
\usepackage{footnote}
\usepackage{tabularx}
\usepackage{hyperref}

\newcommand{\be}{\begin{equation}}
\newcommand{\ee}{\end{equation}}
\newcommand{\bq}{\begin{eqnarray}}
\newcommand{\eq}{\end{eqnarray}}
\newcommand{\ba}{\begin{align}}
\newcommand{\ea}{\end{align}}
\usepackage{slashed}
\usepackage{tabularx}
\usepackage{pifont}
\usepackage[mathscr]{euscript}

\usepackage[latin1]{inputenc}
\usepackage[english]{babel}
\usepackage[T1]{fontenc}
\usepackage{graphicx}
\usepackage{amsmath}
\usepackage{amsfonts,amssymb,ifthen,amsthm}
\usepackage{amsfonts}
\usepackage{amssymb}
\usepackage{subfigure}
\usepackage{multirow}

\begin{document}

\title[Soft Scale Symmetry Breaking with Laguerre Explicit Regulators]{Soft Scale Symmetry Breaking with Laguerre Explicit Regulators}

\date{\today}

\author[]{A. R. Vieira}\email{alexandre.vieira@uftm.edu.br}

\affil[]{\orgdiv{Departamento de Ci\^encias Exatas e Educa\c{c}\~ao}, \orgname{Instituto de Ci\^encias Agr\'arias, Exatas e Biol\'ogicas de Iturama-ICAEBI, Universidade Federal do Tri\^{a}ngulo Mineiro}, \orgaddress{\street{Av. Ant\^onio Baiano, n 150}, \city{Iturama}, \postcode{38280-000}, \state{MG}, \country{Brazil}}}

\abstract{In this work, we search for explicit regulators that do not spuriously break scale invariance starting from conditions established by implicit regularization. There is a family of regulator functions that obeys the conditions and leads to soft breaking terms of the dilatation current. This family of functions is shown to be related to the Laguerre polynomials. We apply this approach by revisiting quadratic divergences in the context of the naturalness problem of the Standard Model and show that the quadratic dependence of the Higgs mass on the cutoff is not required by consistency of scale-symmetry breaking and can be removed within this class of regulators. Applications of these regulators beyond 1-loop order are also
considered.}


\maketitle

\section{Introduction}
\label{s1} 

Symmetries are among the main ingredients to build field theories. Although they are usually taken for 
granted at the classical level, they can be tested in precision experiments, like the Lorentz and CPT ones 
\cite{Kostelecky:2008ts,Colladay:1998fq}, and checked beyond tree-level by computing quantum corrections. It was at first thought that if theories possess symmetries at the classical level, they would keep them at the quantum one as well. This may be true for some symmetries, such as gauge or Lorentz symmetries, but not for others, such as conformal or chiral symmetries. The breaking of classical symmetries when quantum corrections are taken into account is called an anomaly. Some of them are measurable because they are related to physical processes, like the Adler-Bell-Jackiw anomaly (ABJ-anomaly) \cite{Bell, Adler}, related to the pion decay into two photons, and the trace anomaly, related to hadronic processes \cite{Ellis} and to the glue-ball spectrum \cite{Boschi}. The trace of the energy-momentum tensor, which is equal to the dilatation current conservation associated to scale symmetry, is broken at the classical level by massive terms so that massless field theories are scale invariant.  However, using these theories for short distances reveals that scale symmetry is also broken when loop corrections are taken into account \cite{Callan}.

The trace anomaly also manifests itself as a breaking of conformal symmetry in curved space when matter 
fields are embedded in a gravitational background \cite{Duff}.The computation of the energy momentum trace 
performed with different regularization schemes reveals universal pieces of this symmetry breaking \cite
{Duff2}-\cite{IRegTA}. It is sometimes a non-trivial problem to find out if the symmetry breaking was caused 
by the  regularization scheme. From physics point of view, it is interesting to know if the anomaly is an 
artifact of the regularization employed or if it is indeed physical since a regularization scheme can 
spuriously break symmetries of the theory. For instance, we know that cutoff regularization spuriously 
breaks gauge symmetry \cite{Cynolter0} and lattice regularization spuriously breaks Lorentz symmetry 
\cite{Costa}. These spurious symmetry breaking terms can be removed in the renormalization process. However, 
it is necessary to identify if the symmetry breaking term is physical or not with the use of a Ward identity 
or phenomenology. In particular, it has been shown that the implementation of an UV cutoff can explicitly preserve or break fundamental symmetries, such as gauge and Poincaré invariance, depending on its construction \cite{Demir, Demir2}.

In this work, we search for explicit regulators that do not spuriously break scale symmetry. It is well known that scale symmetry is broken by the mass term and due to the renormalization group scale introduced in the renormalization process. Thus, we define a soft scale symmetry breaking regulator as one that does not produce additional breaking terms caused by the use of the regularization scheme itself. We call these 
symmetry breaking terms of scale symmetry as spurious. As a consequence, after the use of the soft scale 
symmetry breaking regulator, it should produce only terms proportional to $m^2$  because they will vanish in 
the limit $m\rightarrow 0$ so that we recover the classical limit, where $m$ is the mass of the particle
in a scale invariant theory in the massless limit. 

The idea of requiring scale symmetry to solve issues with quadratic divergences is not
new \cite{Grange}- \cite{NPB}. These works are usually interested in an investigation of the well
known naturalness problem of the Standard Model (SM) \cite{Susskind, Veltman} and the consequent fine-tuning problem. In reference \cite{Aoki}, the authors argue that quadratic divergences can be absorbed in the definition of a critical surface, based in the Wilsonian renormalization group approach. While in reference
\cite{MPLA}, classical scale invariance is considered as a guideline in models that communicate via a dark sector its scale to the Higgs. Also, in reference \cite{Tavares}, authors show that if the protection to the Higgs mass is attributed to a conformal theory in the ultra-violet (UV), there is an introduction of a new physics scale that keeps the Higgs mass sensitive to large corrections. Related approaches have also explored the possibility that interacting UV fixed points can lead to calculable radiative symmetry breaking in the infrared (IR) \cite{RSB}. Furthermore, the scale invariance in the UV limit can also cause a dynamical breaking that affects the Higgs potential via gauge mediation \cite{NovelHiggs}. 

Spontaneous symmetry breaking (SSB) is not the only way to generate mass. Particle masses can also arise
from radiative corrections \cite{CWM}. If an alternative to the usual SSB is considered, quadratic divergences do not appear \cite{Laporte}. It is worth noting that the interpretation of quadratic divergences as regulator dependent contributions has also been emphasized in the context of dimensional regularization \cite{DREGWIL, Fujikawa} and higher-derivative regularization \cite{Fujikawa}.

In particular, in previous works by some of us \cite{IJTP,AOP}, the existence of regulators that softly break scale symmetry was conjectured within the framework of implicit regularization, based on the physical condition established by the Ward identity of the dilatation current \cite{Bardeen}. Here, we show that such regulators exist and they are related to the Laguerre polynomials, as a consequence of their orthogonality on the interval $[0,\infty[$. Nevertheless, this approach is not necessary if an additional symmetry between bosons and fermions in an extended super-symmetric (SuSy) SM is revealed, unless we are interested in non-soft SuSy explicit breaking \cite{Grisaru2}. Explicit SuSy breaking models are more phenomenological than the SSB ones and although the soft explicit SuSy breaking terms break scale symmetry, quadratic divergences are still absent in these models.

This work is divided as follows: in the next section, we present a short summary of scale symmetry. In 
section \ref{s3}, we present the implicit regularization scheme, show how it can be used to isolate basic 
divergent integrals (BDI) and how quadratic divergences appear in an on-mass-shell renormalization of the 
Higgs propagator. In section \ref{s4}, we discuss how explicit soft scale symmetry breaking regulators can 
be found and how they are related to the Laguerre polynomials. In section \ref{s5}, we discuss how surface 
terms that appear in implicit regularization can be fixed using gauge invariance and how this is compatible with soft scale invariance. In section \ref{s6}, we show how the soft scale symmetry breaking regulators can be understood as a continuous sum of sharp cutoff's of variable radius, each one contributing for the regulated integral with a different weight. In section \ref{s7}, we revisit the naturalness problem of the Standard Model by computing the 1-loop correction to the Higgs mass. Furthermore, we show how to extent
the Laguerre explicit regulators for amplitudes beyond 1-loop in section \ref{s8} and we present conclusions in section \ref{s9}.

\section{Scale symmetry and the dilatation current}
\label{s2}

A theory displays scale invariance if it stays unchanged under the following symmetry transformations,

\be
x'= e^{-\alpha} x
\label{eq0}
\ee

and

\be
\Phi'(x')= e^{\alpha d_{\Phi}} \Phi(x),
\label{eq1}
\ee
where $\alpha$ is a scale parameter, $\Phi$ is a generic field and $d_{\Phi}$ is its scale dimension.

As an example, the kinetic terms of the electromagnetic or spinor fields, $-\frac{1}{4}F^{\mu\nu}F_{\mu\nu}$ 
and $\bar{\psi}\gamma^{\mu}\partial_{\mu}\psi$, respectively, are scale invariant because they possess scale 
dimension equal to 4. In this way, when the volume element $d^4x$ changes  with the use of eq. (\ref{eq0}), 
it compensates the change in the fields making a scale invariant theory. On the other hand, mass terms or 
interaction terms with  dimensionful coupling constants break scale symmetry at the classical level because 
they do not transform under scale symmetry.

The Noether current related to the transformations presented in eqs. (\ref{eq0}) and (\ref{eq1}) is the 
dilatation current $\mathcal{J}^{\mu}=\Pi^{\mu}_id_{\Phi_i}\Phi_i+x_{\nu}\Theta^{\mu\nu}$, 
where $\Pi^{\mu}_i=\frac{\partial \mathcal{L}}{\partial (\partial_{\mu}\Phi_i)}$ is the canonical conjugate 
momentum and $\Theta^{\mu\nu}$ is the standard energy momentum tensor given by 
$\Theta^{\mu\nu}=\Pi^{\mu}_i\partial^{\nu}\Phi_i-g_{\mu\nu}\mathcal{L}$, where $i$ stands for the $i-th$ 
field. We use the canonical energy-momentum tensor since our discussion concerns scale invariance and the dilatation current. Improvement terms do not modify the mass-dependent quadratic contribution, that we
consider throughout this work.

Besides the dimensionful terms, like mass terms in the Lagrangian, the renormalization procedure itself also breaks scale symmetry because it introduces a renormalization group scale dependence in the couplings. This 
scale dependence is introduced in the stage of any explicit regularization. For instance, in dimensional 
regularization \cite{DR,Bollini}, we are forced to introduce a scale $\mu$ to keep the correct energy dimension of the divergent integral when we change the space-time dimension to $D$. However, regardless of the regularization or the renormalization adopted, this effect is known as the anomalous breaking of the scale symmetry and it is physically expected since the same theory considered at larger scale is characterized by a different value of the renormalized coupling constant.

As an example, let us consider the simplest theory, a massive real scalar theory with quartic interaction:
\begin{equation}
\mathcal{L}=\frac{1}{2}\partial_{\mu}\phi\partial^{\mu}\phi-\frac{1}{2}m^2\phi^2-\frac{\lambda}{4!}\phi^4,
\label{eqLS}
\end{equation}
where $m$ and $\lambda$ are the scalar mass and self-interacting coupling, respectively.

Considering the Lagrangian in eq. (\ref{eqLS}), we have $\Pi^{\mu}=\frac{\partial 
\mathcal{L}}{\partial(\partial_{\mu}\phi)}=\partial^{\mu}\phi$ and 
$\Theta^{\mu\nu}=\partial^{\mu}\phi\partial^{\nu}\phi-g^{\mu\nu}\mathcal{L}$. Consequently, one finds 
$\Theta^{\mu}_{\ \ \mu}=m^2\phi^2$ after the use of the equations of motion and the dropping of total 
derivatives. This is only the classical breaking. However, the quantum corrections also break scale symmetry. There is a correction to the mass term due to the self energy of the scalar and the renormalization process introduces a scale, that necessarily breaks scale symmetry, such that the full breaking of the trace of the energy momentum tensor is given by:
\begin{equation}
\Theta^{\mu}_{\ \ \mu}=m^2_r\phi^2+\beta_{\lambda}\frac{\partial \mathcal{L}}{\partial \lambda},
\end{equation}
where $\beta_{\lambda}$ is the beta function of the coupling and $m^2_r=m^2+\delta m^2$ is the renormalized 
on-shell mass.

All regularization and renormalization schemes produce the expected quantum breaking term proportional to 
the $\beta$ function. However, if a regularization scheme spuriously breaks scale symmetry, we do not recover $\Theta^{\mu}_{\ \ \mu}\rightarrow 0$ when we take the limits $m\rightarrow 0$ and 
$\beta_{\lambda}\rightarrow 0$. As a consequence, the quantum correction of the mass $\delta m^2$ should be 
proportional to $m^2$.

\section{Implicit Regularization and On-Mass-Shell Renormalization}
\label{s3}

We apply the implicit regularization framework \cite{Orimar0}-\cite{Adriano3} to treat the divergent 
integrals. In this scheme,  an implicit regulator is assumed so that it makes sense to perform algebraic 
operations with the integrands. If such regulator exists, it should obey the property $\lim_{\Lambda \to 
\infty}R(\Lambda^2,k^2)=1$ such that:
\begin{align}
\int \frac{d^4k}{(2\pi)^4}f(k) &\rightarrow\int \frac{d^4k}{(2\pi)^4}f(k)\lim_{\Lambda \to \infty}R(
\Lambda^2,k^2) \nonumber\\
&= \int^{\Lambda} \frac{d^4k}{(2\pi)^4}f(k),
\end{align}
where $\int \frac{d^4k}{(2\pi)^4}f(k)$ is a divergent integral and the index $\Lambda$
means that an implicit regulator is assumed. However, it is not necessarily a sharp cutoff.

The regulator function $R(\Lambda^2,k^2)$ besides having a well defined limit, must be dimensionless and 
even in the integrated momentum due to Lorentz invariance. For the
purposes of this work it also should not spuriously break scale symmetry.

When the implicit regulator is assumed, it makes mathematical sense to perform operations
with the integrand. It is possible to use, for example, the following identity to separate UV 
basic divergent integrals (BDI) from the finite part:

\begin{align}
&\int_k \frac{1}{(k+p)^2-m^2}=\int_k\frac{1}{k^2-m^2}\nonumber\\
&-\int_k\frac{(p^2+2p\cdot k)}{(k^2-m^2)[(k+p)^2-m^2]},
\label{2.1}
\end{align}
where $\int_k\equiv\int^\Lambda\frac{d^4 k}{(2\pi)^4}$. These BDI's are defined as follows:
\begin{equation}
I^{\mu_1 \cdots \mu_{2n}}_{log}(m^2)\equiv \int_k \frac{k^{\mu_1}\cdots k^{\mu_{2n}}}{(k^2-m^2)^{2+n}},
\label{eqBDI1}
\end{equation}
and
\begin{equation}
I^{\mu_1 \cdots \mu_{2n}}_{quad}(m^2)\equiv \int_k \frac{k^{\mu_1}\cdots k^{\mu_{2n}}}{(k^2-m^2)^{1+n}}.
\label{eqBDI2}
\end{equation}

The basic divergences with Lorentz indices can be combined as differences between integrals with the same 
superficial degree of divergence, according to the equations below, which define surface terms \footnote{The 
Lorentz indices between brackets stand for permutations, i.e. 
$A^{\{\alpha_1\cdots\alpha_n\}}B^{\{\beta_1\cdots\beta_n\}}=A^{\alpha_1\cdots\alpha_{n}}B^{\beta_1\cdots\beta
_n}$ + sum over permutations between the two sets of indices $\alpha_1\cdots\alpha_{n}$ and 
$\beta_1\cdots\beta_n$}:
\begin{align}
&\Upsilon^{\mu \nu}_{2w}= g^{\mu \nu}I_{2w}(m^2)-2(2-w)I^{\mu \nu}_{2w}(m^2) \nonumber\\
&\equiv \upsilon_{2w}g^{\mu \nu},
\label{dif1}\\
\nonumber\\
&\Xi^{\mu \nu \alpha \beta}_{2w}=  g^{\{ \mu \nu} g^{ \alpha \beta \}}I_{2w}(m^2)
 \nonumber\\& - 4(3-w)(2-w)I^{\mu \nu \alpha \beta }_{2w}(m^2) \equiv  \xi_{2w}g^{\{ \mu \nu} g^{ \alpha 
\beta \}}.
\label{dif2}
\end{align}

In the expressions above, $2w$ is the degree of divergence of the integrals, and,  for the sake of brevity, 
we substitute the subscripts $log$ and $quad$ by $0$ and $2$, respectively. Surface terms can be 
conveniently written as integrals of total derivatives, namely
\be
\upsilon_{2w}g^{\mu \nu}= \int_k\frac{\partial}{\partial k_{\nu}}\frac{k^{\mu}}{(k^2-m^2)^{2-w}},
\label{ts1}
\ee
and
\be
(\xi_{2w}-v_{2w})g^{\{ \mu \nu} g^{ \alpha \beta \}}= \int_k\frac{\partial}{\partial k_{\nu}}\frac{2(2-w)k^{
\mu} k^{\alpha} k^{
\beta}}{(k^2-m^2)^{3-w}}.
\label{ts2}
\ee

Equation (\ref{2.1}) is applied in a recursive way till the BDI's are separated from the finite content of 
the amplitude. Also, we see that the surface terms defined in equations (\ref{dif1})-(\ref{dif2}) are 
undetermined because they  are differences between two divergent quantities. Each regularization scheme 
gives a different value for these terms. However, as physics should not depend on the regularization, we 
leave these terms to be arbitrary until the end of the calculation, fixing them by symmetry constraints, 
like a Ward identity, or phenomenology \cite{Jackiw2}.

\begin{figure}[htb]
\centering
\includegraphics[trim= 0mm 20mm 0mm 20mm, scale=0.7]{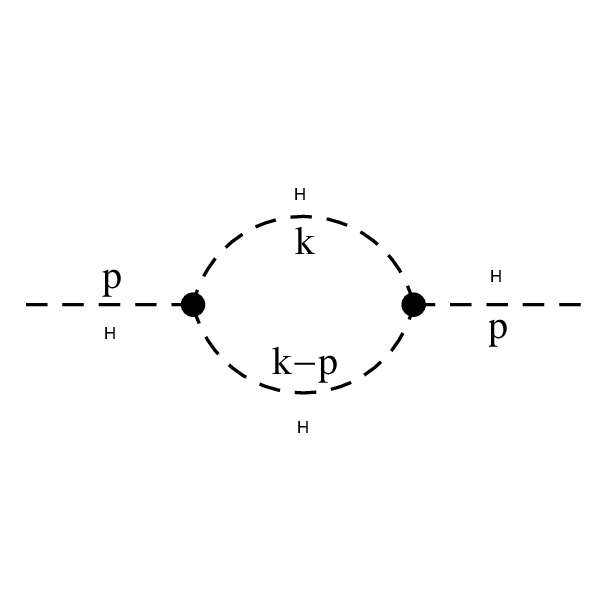}
\caption{1-loop correction to the Higgs propagator.}
\label{fig0}
\end{figure}

The prescription above is not the usual regularization procedure and it is particularly useful in the 
computation of anomalies, when we need to know if the quantum breaking of a symmetry really occurs or if it 
is a spurious symmetry breaking caused by the regulator. As an example, let us consider the 1-loop 
correction to the Higgs propagator presented in figure \ref{fig0}:

\begin{align}
&\int^{\Lambda}\frac{d^4k}{(2\pi)^4}\frac{1}{(k^2-m^2)[(k-p)^2-m^2]}=I_{log}(m^2)\nonumber\\
&-\int^{\Lambda}\frac{d^4k}{(2\pi)^4}\frac{p^2-2p\cdot k}{(k^2-m^2)^2[(k-p)^2-m^2]}\nonumber\\
&=I_{log}(m^2)-b Z_0(p^2,m^2),
\label{eq1LH}
\end{align}
where $b\equiv\frac{i}{(4\pi)^2}$, $Z_0(p^2,m^2)\equiv Z_0=\int^1_0dx \ln\left[\frac{m^2-p^2x(1-x)}{m^2}\right]$ and the identity (\ref{2.1}) was applied once. The finite content can be computed analytically by the usual Feynman parametrization. 

All the regularized integrals found in the amplitudes of this work are presented in the appendix. Actually, the use of the recursive relation of eq.(\ref{2.1}) makes the computation of the finite content of the amplitude more involved because the number of $k$'s in the numerator increases, especially if the starting integral has a high superficial divergent degree. An alternative to avoid this issue is proposed in \cite{Bruno}, where Feynman parametrization is applied prior the isolation in BDI's. This alternative aims to ease the computation of the finite content of the amplitudes. However, when it is applied, it assumes momentum routing invariance of the divergent amplitudes. Thus, the version of implicit regularization that we adopt in this manuscript is the original one \cite{Orimar0}, where no {\it ad hoc} assumptions are made besides the existence of the implicit regulator. This avoids additional surface terms generated when doing
shifts in the integrated momentum of the divergent integrals in the Feynman parametrization process.

\begin{figure}[htb]
\centering
\includegraphics[trim=0mm 0mm 0mm 0mm,clip, scale=0.8]{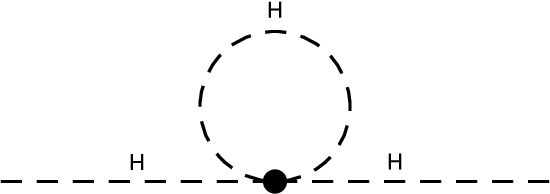}
\caption{Correction to the Higgs propagator with a Higgs boson loop.}
\label{fig1a}
\end{figure}

\begin{figure}[htb]
\center
\includegraphics[trim=0mm 0mm 0mm 0mm,scale=0.7]{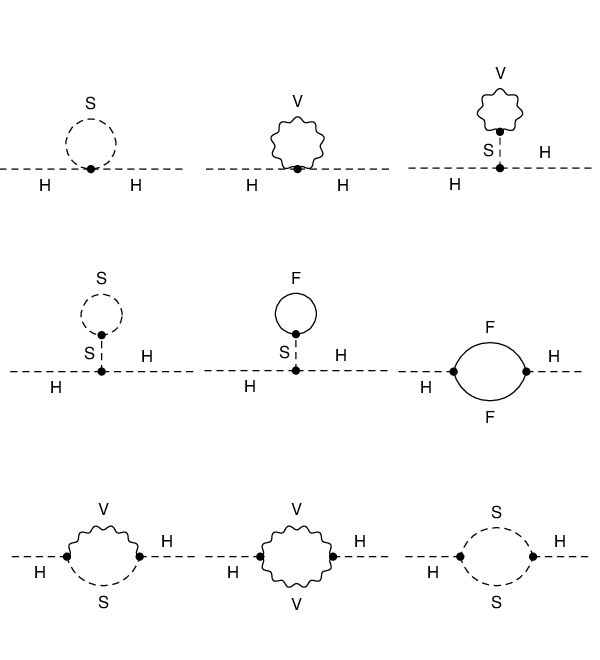}
\caption{1-Loop correction to the Higgs propagator in the Landau gauge. S, V and F stands for scalar, vector 
boson and fermionic fields, respectively.}
\label{figDiag}
\end{figure}

Some 1-loop diagrams of the Higgs propagator directly produce quadratic divergent BDI's that do not 
have to be isolated with the use of implicit regularization, like the one presented in figure \ref{fig1a}. 
The corresponding amplitude of this diagram is
\begin{equation}
\frac{3}{2}\lambda I_{quad}(m^2)=\frac{3m^2}{4\nu^2} I_{quad}(m^2),
\label{eqQC}
\end{equation}  
where $\lambda$ is the Higgs self-coupling which is related to the Higgs mass using the
eq. $\lambda=\frac{m^2}{2\nu^2}$ being $\nu$ the vacuum expectation value of the
Higgs field. The number $3/2$ is a symmetry factor. They appear in most of the 1-loop
diagrams for the Higgs propagator whose form is presented in figure \ref{figDiag} in the
Landau gauge.

We see in eq. (\ref{eqQC}) that the Higgs sector is protected against a hard breaking of scale symmetry 
because of the self coupling that automatically cancels the quadratic divergence in the limit $m\rightarrow 
0$. This is of course independent of regularization or renormalization schemes. It comes simply from the 
Lagrangian of the Higgs sector:
\begin{eqnarray}
\mathcal{L}_H(x)&=&[D^{\mu}\Phi(x)]^{\dagger}[D_{\mu}\Phi(x)]-\mu^2\Phi(x)^{\dagger}\Phi(x)
\nonumber \\
&-&\lambda[\Phi(x)^{\dagger}\Phi(x)]^2,
\label{eq13}
\end{eqnarray}
where $\Phi(x)=\begin{pmatrix}\phi^+\\\phi^0\end{pmatrix}$ is the Higgs field, $D_{\mu}$ is the covariant 
derivative that couples it with the gauge fields and $\mu^2<0$. 

As in the real scalar field theory, the Higgs mass term breaks the classical conservation law because it is 
the only one in the Lagrangian which does not possess scale dimension equal to four, {\it i. e.} the mass  
did not transform according to the scale transformations. Therefore,
\begin{equation}
\Theta^{\mu}_{\ \ \mu}=m^2\Phi(x)^{\dagger}\Phi(x),
\label{eq15}
\end{equation}
where $m^2=-2\mu^2$ is the tree-level Higgs mass.

Also, as presented previously in section \ref{s2}, since quantum corrections make the couplings depend on  
the renormalization group scale, the Lagrangian (\ref{eq13}) changes due to the scale transformation
\be
x'=e^{-\alpha}x \rightarrow \Lambda'=e^{\alpha}\Lambda,
\ee
which leads to
\begin{align}
&& \delta\mathcal{L}_H(m_r(\Lambda),\lambda(\Lambda))= \nonumber \\
&& \alpha\{m^2_r(\Lambda) \gamma \Phi(x)^{\dagger}\Phi(x)+\beta_{\lambda}[\Phi(x)^{\dagger}\Phi(x)]^2\},
\label{eq16}
\end{align}
where $m^2_r(\Lambda)$ and $\lambda(\Lambda)$ are the renormalized Higgs mass and self-coupling, 
respectively, and
\begin{equation}
 \gamma=\frac{\Lambda^2}{m_r^2(\Lambda^2)}\frac{\partial m_r^2(\Lambda^2)}{\partial \Lambda^2}
\label{eq17}
\end{equation}
is the renormalization group gamma function. Hence, the complete violation of the dilatation current
considering the Higgs sector, due to the mass term and the quantum corrections, is given by

\begin{align}
\Theta^{\mu}_{\ \ \mu}&=(m^2+m^2_r(\Lambda) \gamma) \Phi(x)^{\dagger}\Phi(x) \nonumber\\
&-\beta_{\lambda}[\Phi(x)^{\dagger}
\Phi(x)]^2.
\label{eq18}
\end{align}

\begin{figure}[htb]
\centering
\includegraphics[trim=0mm 35mm 0mm 0mm,clip, scale=0.8]{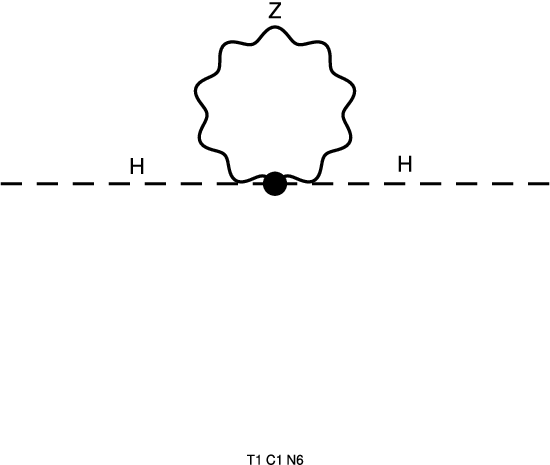}
\caption{Correction to the Higgs propagator with a $Z$-boson loop.}
\label{fig1b}
\end{figure}

We do not have to consider in eq. (\ref{eq18}) the complete version of the trace $\Theta^{\mu}_{\ \ \mu}$
in the full SM because only the Higgs sector is affected by the quadratic divergences. However, even in the
Higgs sector, the problem of quadratic divergences appears due to the Higgs interaction with other massive particles. For instance, the 1-loop diagram with a massive gauge boson $Z$, presented in figure \ref{fig1b}, whose amplitude is given by
\begin{align}
&\frac{i g^2}{4 \cos \theta_W} g_{\mu\nu}\int \frac{d^4k}{(2\pi^4)}\frac{-i}{k^2-m_Z^2}\left(g^{\mu\nu}-\frac
{k^{\mu}k^{\nu}}{k^2}\right) \nonumber\\ 
&=\frac{3m^2_Z}{\nu^2}I_{quad}(m^2_Z),
\label{eqDZ}
\end{align}
where $m_Z$ is the $Z$-boson mass and $\nu$ is the vacuum expectation value.

Now, in order to renormalize the propagator, we consider the on-mass-shell (OMS) renormalization with the 
use of the equation below:
\begin{equation}
\frac{1}{A-B}=\frac{1}{A}+\frac{1}{A}B\frac{1}{A}+\frac{1}{A}B\frac{1}{A}B\frac{1}{A}+...
\end{equation}
which implies that
\begin{align}
&\frac{i}{p^2-m^2_r}=\frac{i}{p^2-m^2}\nonumber\\
&+\frac{i}{p^2-m^2}\frac{-3im_Z^2}{\nu^2}I_{quad}(m^2_Z)\frac{i}{p^2-m^2}+...,
\label{eqOMS}
\end{align}
where $m_r$ stands for the renormalized Higgs mass and the ellipsis stands for the
contributions of the other diagrams. In this case, the mass is considered as the pole of the propagator. The 
optical theorem states that the contribution to the tree-level mass comes from the real part of the 1-loop 
corrections to the tree-level propagator.

With the use of eq. (\ref{eqOMS}) we find out that this 1-loop correction breaks scale
symmetry because it is proportional to the $Z$-boson mass, unlike the one coming from 
the Higgs self coupling:
\begin{equation}
m^2_r=m^2-\frac{3im_Z^2}{\nu^2}I_{quad}(m^2_Z)+...,
\label{eqRM}
\end{equation}
where the ellipsis stands for the finite and logarithmic divergent corrections and the other 1-loop diagrams 
of the Higgs propagator.

Based on the corrections like the ones presented in eq. (\ref{eqRM}), in the next section, we look for 
regulators that will softly break scale symmetry.

\section{Soft scale symmetry breaking regulators}\label{s4}

The BDI's of the previous sections would explicitly depend on the regulator if it is 
specified. Based on the condition $\lim_{\Lambda \to \infty}R(\Lambda^2,k^2)=1$, it is easy to check that 
$R(\Lambda^2,k^2)=e^{k^2/\Lambda^2}$ meets this criterion but it is of course not the only possible regulator function. In particular, choices like this one do not necessarily mean that the regulator is invariant under the desired symmetry. Besides this criterion, we require the additional conditions that should be obeyed by any regularization scheme:
\begin{align}
&\frac{\partial I_{quad}(m^2)}{\partial m^2}=I_{log}(m^2),\nonumber\\
&\frac{\partial I_{log}(m^2)}{\partial m^2}=\frac{-b}{m^2},
\label{eqDIFFEQ}
\end{align}
where $b\equiv\frac{i}{(4\pi)^2}$. Notice that the derivative is partial because the BDI's depend on the
implicit regulator $\Lambda$.

It is straightforward to check that a general parametrization that obeys eqs. (\ref{eqDIFFEQ}) can be 
written as
\be
I_{log}(m^2)=b\left( \ln\frac{\Lambda^2}{m^2}+c_1\right)\nonumber\\
\label{eqPARAM0}
\ee
and
\be
I_{quad}(m^2)=b\left( c_2\Lambda^2 +(1+c_1)m^2+m^2\ln\frac{\Lambda^2}{m^2}\right)
\label{eqPARAM}
\ee

In scalar theories, we know that quantum corrections should break the trace of the energy momentum tensor by 
terms proportional to $m^2$. This information already restricts
what kind of explicit regulators we should have, those for which $c_2=0$. In this way, we recover the 
classical limit when we take $m\rightarrow 0$ and $\beta\rightarrow 0$, as discussed previously in section 
\ref{s2}. We also know from eq. (\ref{eqBDI2}) that $I_{quad}(m^2)=\int_k\frac{1}{k^2-m^2}$ and this can be 
written in the euclidean space after Wick rotation as:
\begin{align}
&I_{quad}(m^2)=\int_k\frac{1}{k^2-m^2}=-i\int_0^{\infty} \frac{dk_E}{8\pi^2} \frac{k^3_E}{k^2_E+m^2}
\nonumber\\
&\rightarrow -\frac{i}{8\pi^2}\int_0^{\infty} dk_E \frac{k^3_E R(k^2_E,\Lambda^2)}{k^2_E+m^2}\nonumber\\
&=-b\Lambda^2\int_0^{\infty}dx \frac{xR(x)}{x+a},
\end{align}
where $x=\frac{k^2_E}{\Lambda^2}$ and $a=\frac{m^2}{\Lambda^2}$.

Now, it is possible to use the relation $\frac{x}{x+a}=1-\frac{a}{x+a}$ to split the integrand of the 
rewritten term in the previous equation so that we have:
\begin{align}
I_{quad}(m^2)&=-\frac{i\Lambda^2}{16\pi^2}\int_0^{\infty}dx R(x)\nonumber\\
&+\frac{i m^2}{16\pi^2}\int_0^{\infty}dx \frac
{R(x)}{x+a},
\label{eqSPLIT}
\end{align} 
where we can directly identify $c_2=-\int_0^{\infty}dx R(x)$ using eq. (\ref{eqPARAM}).

We see that there is no problem with scale symmetry breaking in the second term of eq. (\ref{eqSPLIT}) 
because it scales as $m^2$.  However, if the explicit regulator is $ 
R(\Lambda^2,k^2_E)=e^{-k^2_E/\Lambda^2}=e^{-x}$, it is not possible to have $\int_0^{\infty}dx R(x)=0$ so 
that the whole integral softly breaks scale symmetry. In this way, not all the regulators that are according to implicit regularization requirements, {\it i. e.} $\lim_{\Lambda \to \infty}R(\Lambda^2,k^2)=1$, necessarily comply with soft break of scale symmetry.  Nevertheless, we do not have to go far to find out that if the regulator function is, for instance, $R(x)=(1-x)e^{-x}$, we do have $c_2=0$. Now, returning to the regulated quadratic BDI, we can easily write it as

\begin{equation}
I_{quad}(m^2)=\int\frac{d^4k}{(2\pi)^4}\frac{\left(1+\frac{k^2}{\Lambda^2}\right)e^{k^2/\Lambda^2}}{k^2-m^2},
\end{equation}
from which we can easily check that the regulator is according to eqs. (\ref{eqDIFFEQ}) in the limit 
$\Lambda\rightarrow \infty$. There is also a change of sign due to the Wick rotation, 
{\it i. e.} $k^2\rightarrow -k^2_E$.

There is actually a family of functions $R_n(x)=P_n(x)e^{-x}$ such that $P_n(0)=1$, with the property 
$\int_0^{\infty}dx R_n(x)=0$. For example, all the Laguerre polynomials have this property because 
$L_n(0)=1$. However, the one for $n=0$ is the only one that breaks scale symmetry. So that we can choose 
$P_n(x)=L_n(x)$ for $n\neq0$ as can be easily checked for any $n$ of the table \ref{tab00}. This is expected 
since $R_n(x)$ is related to the generator of the Laguerre polynomials using the Rodrigues formula:
 $R_n(x)=L_n(x)e^{-x}=\frac{1}{n!}\frac{d^n}{dx^n}(x^ne^{-x})$. Therefore, we find out that
\begin{align}
c_2&=-\int^{\infty}_0 L_n(x)e^{-x} dx\nonumber\\
&=-\frac{1}{n!}\int^{\infty}_0 \frac{d^n}{dx^n}(x^ne^{-x}) dx,
\end{align}
which is null for all $n>0$. This is a consequence of the differential structure of the Laguerre polynomials.

\begin{table}
\centering
 \caption{Laguerre polynomials and the corresponding value for the coefficient $c_2$.}
 \begin{tabular}{ccc}
  \hline
  $Degree$& Polynomial&Coefficient $c_2$ \\
  \hline
  $n=0$ & $1$& $-\int^{\infty}_0 L_0(x)e^{-x}=-1$ \\
  \hline
   $n=1$& $1-x$& $-\int^{\infty}_0 L_1(x)e^{-x}=0$ \\
  \hline
  $n=2$&  $\frac{x^2}{2}-2x+1$& $-\int^{\infty}_0 L_2(x)e^{-x}=0$ \\
  \hline
  $n=3$&  $-\frac{x^3}{6}+...+1$& $-\int^{\infty}_0 L_3(x)e^{-x}=0$ \\
  \hline
  ...&...&...\\
  \hline
  $n$&$\frac{e^{x}}{n!}\frac{d^n}{dx^n}(x^ne^{-x})$&$-\int^{\infty}_0 L_n(x)e^{-x}=0$\\
  \hline 
 \end{tabular}
\label{tab00} 
\end{table}

We understand now why the $n=0$ is the only one that does not make the coefficient $c_2$
equal to zero. For all the other cases, we can write $c_2$ as an integral of a total derivative in the 
dimensionless radial space:
\begin{align}
c_2&=-\frac{1}{n!}\int^{\infty}_0 \frac{d}{dx}\left(\frac{d^{n-1}}{dx^{n-1}}(x^ne^{-x})\right) dx\nonumber\\
   &=-\frac{1}{n!}\left(\frac{d^{n-1}}{dx^{n-1}}(x^ne^{-x})\right) \Big|^{\infty}_0=0.
\label{eqSTC2}
\end{align}

In this way, a class of explicit regulators for the quadratic basic divergent integral that does not hard break scale symmetry can be written as
\begin{equation}
I^n_{quad}(m^2)=\int\frac{d^4k}{(2\pi)^4}\frac{L_n(-\frac{k^2}{\Lambda^2})e^{k^2/\Lambda^2}}{k^2-m^2}, 
\text{for}\ n\neq 0.
\end{equation} 

The other coefficient, $c_1$, does not lead to hard scale symmetry breaking because it is multiplied by 
$m^2$. However, we are going to write it for the sake of completeness. By comparing
eqs. (\ref{eqPARAM}) and (\ref{eqSPLIT}), we can write:
\begin{equation}
c_1=\int^{\infty}_0 dx \frac{R(x)}{x+a}-\ln\frac{\Lambda^2}{m^2}-1,
\end{equation}
where $a=\frac{m^2}{\Lambda^2}$ as before. 

If $R(x)=(1-x)e^{-x}$, it is easy to see that $\ln \frac{\Lambda^2}{m^2}$ vanishes and that $c_1=-2-\gamma_E$
in the limit $\Lambda\rightarrow \infty$, where $\gamma_E$ is the Euler-Mascheroni constant because
\begin{equation}
\int^{\infty}_0 dx \frac{R(x)}{x+a}=-1+(1+a)e^a \Gamma(0,a),
\end{equation}
where $\Gamma(0,a)=\int^{\infty}_a \frac{e^{-t}}{t}dt$ is the upper incomplete gamma function, which implies 
that $\Gamma(0,a)=-\gamma_E-\ln a+a +...$ for $a<<1$.

The situation is similar for $R_n(x)=L_n(x)e^{-x}$, {\it i. e.} the term $\ln \frac{\Lambda^2}{m^2}$ still 
cancels but the value of $c_1$ changes. In any case, all these terms vanish in the conformal limit 
$m\rightarrow 0$, which means that the regulators softly break scale invariance for all $n\neq 0$.

It is also easy to check that the logarithmic BDI $I_{log}(m^2)$ can be regulated by this
family of regulators:
\begin{align}
&I_{log}(m^2)=-\frac{i}{8\pi^2}\int_0^{\infty} dk_E \frac{k^3_E}{(k^2_E+m^2)^2}\nonumber\\
&\rightarrow \frac{i}{8\pi^2}\int_0^{\infty} dk_E \frac{k^3_E R(k^2_E,\Lambda^2)}{(k^2_E+m^2)^2}\nonumber\\
&=b\int_0^{\infty}dx \frac{xR(x)}{(x+a)^2},
\label{BDI1L}
\end{align}
from which we obtain the logarithmic dependence expanding for $a<<1$:
\begin{align}
\int_0^{\infty}dx \frac{xR(x)}{(x+a)^2}=
\left(\ln\frac{\Lambda^2}{m^2}-2-\gamma_E \right).
\label{eqRL}
\end{align}

In the equation above, we chose to regulate the logarithmic divergence with the same
Laguerre polynomial just to obtain the same coefficient value for $c_1$. However, in this case, there is no 
problem if the regulator is $R_0(x)=L_0(x)e^{-x}$ and the soft scale symmetry breaking regulators are
good for logarithmic divergences considering all the Laguerre polynomials. Eqs. (\ref{eqSPLIT}) and (\ref
{eqRL}) belongs to the step where the renormalization group scale is introduced, $\Lambda$ maps the 
running of the coupling and masses and its maximum value is the cutoff or the limit of validity of the 
theory. 

There are actually other ways to understand why the coefficients of the quadratic divergences, $c_2$, vanish except for $n=0$. We only have to use the orthogonality of the Laguerre polynomials $\int^{\infty}_0 L_i(x) L_j(x) e^{-x}dx=\delta_{ij}$ for $j=0$, so that
\begin{align}
\int^{\infty}_0 L_0(x) L_i(x) e^{-x}dx&=\delta_{i0}\nonumber\\
&\Rightarrow c_2=-<L_0,L_i>.
\end{align}

Besides the mathematical reasoning, from the physical point of view, the condition $P_n(0)=L_n(0)=1$ has the important role of not altering the infrared (IR) regime, while $e^{-x}$ suppresses the UV regime. This is particularly useful for regularizations in the high-energy limit.

It is interesting to notice that a null $c_2$ should not be interpreted as resulting from a fine-tuning. It comes from a generalization of the idea of a hard cutoff, as we are going to present
in section \ref{s6}. The coefficient $c_1$ depends on the Laguerre regularization, while $c_2$ is null for 
all Laguerre polynomials except for $n= 0$, which follows from the requirement that the regulator 
introduces no additional hard breaking of scale symmetry and it is a consequence of the Laguerre polynomials
orthogonality. On the other hand, the coefficient $c_1$ depends on the degree of the Laguerre polynomial chosen because for none of them scale symmetry is broken by a spurious term.

We can check using eq. (\ref{eqPARAM0}) that we get again $c_1=-2-\gamma_E$ for the Laguerre polynomial of degree 1 in eq. (\ref{eqRL}). Thus, we can establish an algorithm to apply the soft scale symmetry breaking regulators in a divergent amplitude:
\begin{enumerate}
    \item Perform all the Clifford, Dirac and Lorentz algebra of the amplitudes.
    \item Apply the recursive relation of implicit regularization in order to isolate the BDI's.
    \item Reduce the BDI's with indices in BDI's without indices using eqs. (\ref{dif1}) and (\ref{dif2}).
    \item Use the scale relation to write all the BDI's as a function of only one mass (this step will be discussed
in section \ref{s7} where we consider loops of the Higgs propagator that have different massive particles).
    \item Choose one of the regulators $R_n(x)$ ($n\neq 0$ if the divergence is quadratic) and fix the 
coefficients $c_2$ and $c_1$.    
\end{enumerate}

As we are going to present in the next section, if step 1 is not performed prior to regularization, it 
may appear extra surface terms or alter the divergent and the finite contents of the amplitudes. If the 
steps 2 to 4 are skipped and the explicit soft breaking regulators are applied in integrals dependent on the
external momenta, it is necessary to use Feynman parametrization and symmetric integration in order to get 
rid of the Lorentz indices of integrated momenta. This also makes the integrals harder to compute than
when they are written in terms of BDI and the result might coincide with the one obtained when steps $1\ to\ 5$ are taken depending on the superficial divergent degree of the amplitude. 

For example, if we use Feynman parametrization, we are assuming momentum routing invariance of the amplitude when we apply the shift in the integrated momentum. This is true for finite pieces and logarithmic divergences but not for amplitudes with higher superficial divergent degree. In order to see this, let us consider again the example of eq. (\ref{eq1LH}). After Feynman parameterization, we find:
\begin{align}
&\int_k \frac{1}{(k^2-m^2)[(k-p)^2-m^2]}=\int^1_0 dz\int_l \frac{1}{(l^2-\Delta )^2}\nonumber\\
&\rightarrow \int^1_0 dz \int_l \frac{(1+l^2/\Lambda^2 ) e^{l^2/\Lambda^2 }}{(l^2-\Delta )^2}\nonumber\\
&=b\left(\ln \frac{\Lambda^2}{m^2}- Z_0-2-\gamma_E  \right),
\end{align}
which is the same result obtained if we use eq. (\ref{eqRL}) in eq. (\ref{eq1LH}).

However, the results from both approaches, $1\ to\ 5$ and $1\ then\ 5$, matches because the amplitude is
logarithmic divergent. In this case, there is no problem in doing the shift $k\rightarrow l+pz$ after Feynman
parametrization because logarithmic divergent integrals are momentum routing invariant. On the other hand, if the superficial divergent degree of the amplitudes was higher, the shift would produce additional surface terms that after evaluated in an explicit regularization (Laguerre or not) would make the final result different. For that reason, we adopt the original version of implicit regularization \cite{Orimar0} in order to separate the divergent content of the amplitude in BDI's before prior computation of any integral.

It is also important to notice that this regularization prescription intends to avoid a hard breaking of
scale symmetry introduced spuriously by the regulator. However, if there is a physical breaking of scale symmetry introduced by a heavy particle it has nothing to do with the regularization scheme and it might
break scale symmetry of the Higgs sector when correcting the Higgs mass at the quantum level. The models that can introduce a physical breaking of scale symmetry do not have to be SuSy ones necessarily. For instance, a massive heavy fermionic field that interacts with the Higgs is considered as a sterile neutrino field in some seesaw models for neutrinos \cite{SeeSawN}:
\begin{equation}
\mathcal{L}\supset -y_{\nu}\bar{L}\tilde{H}\nu_R-\frac{1}{2}M_N\bar{\nu^c}_R \nu_R  ,
\end{equation}
which may imply in corrections to the Higgs mass proportional to $M^2_N$ or $M^2_N\ln\frac{\Lambda}{M_N}$, with $M_N>>m$.

\section{Discussion on surface terms and gauge invariance of QED and the Higgs sector}
\label{s5}

We can apply the implicit regularization scheme, presented in section \ref{s2}, to compute the vacuum 
polarization tensor of QED and find conditions for surface terms required by gauge symmetry. The procedure involves building the amplitude with Feynman rules, taking the trace, regularizing the integrals and then require that the gauge Ward identity be fulfilled. The approach of using a symmetry to
fix arbitrary regularization dependent terms was first proposed in \cite{Jackiw2}. The regularized vacuum 
polarization tensor is well-known and its computation in implicit regularization is given by:

\begin{align}
\centering
i\Pi^{\mu\nu}(p)&=\frac{4}{3}e^2(p^2g^{\mu\nu}-p^{\mu}p^{\nu})I_{log}(m^2_e)\nonumber\\
&-4e^2\upsilon_2g^{\mu\nu}+\frac{4}{3}e^2(p^2g^{\mu\nu}-p^{\mu}p^{\nu})\upsilon_0 \nonumber\\
&-\frac{4}{3}e^2(p^2g^{\mu\nu}+2p^{\mu}p^{\nu})(\xi_0-2\upsilon_0)\nonumber\\
&-\frac{i}{2\pi^2}e^2(p^2g^{\mu\nu}-p^{\mu}p^{\nu})(Z_1-Z_2),
\label{eqpi}
\end{align}
where $m_e$ is the electron mass, $p$ is its external momentum, $Z_n=\int_0^1 dx x^n\ln \left(\frac{\Delta(x)}{m^2_e}\right)$ and $\Delta(x)=m^2_e-p^2 x(1-x)$.

Notice that if we require gauge invariance using the Ward identity $p_{\mu}\Pi^{\mu\nu}(p)=0$, we find that 
the quadratic surface term $\upsilon_2$ must be zero and that the logarithmic surface terms must obey the 
relation $\xi_0=2\upsilon_0$. The finite and the divergent pieces of eq. (\ref{eqpi}) are gauge invariant as 
expected and so is the remaining surface term $\upsilon_0$.

The Higgs propagator is affected by some 1-loop diagrams with gauge boson mediators.
The prescription in implicit regularization is to perform the group algebra before any regularization, 
including the implicit one, because if this is not done, we can generate
additional surface terms or even logarithmic divergences. As an example, we consider
the 1-loop diagram with a $Z-$boson mediator. They may lead to quadratic divergent integrals that are linked 
to the arbitrary surface term $\upsilon_2$ if the Lorentz algebra is not performed prior the regularization. Let us consider again the diagram presented in figure \ref{fig1b}. The Feynman rules lead to the following amplitude:

\begin{align}
\centering
&\frac{i g^2}{8 \cos \theta_W} g_{\mu\nu}\int \frac{d^4k}{(2\pi)^4}\frac{-i}{k^2-m_Z^2}\left(g^{\mu\nu}-\frac{k^{\mu}k^{\nu}}{k^2}\right)\nonumber\\
&=\frac{m^2_Z}{2\nu^2} g_{\mu\nu}\int \frac{d^4k}{(2\pi)^4}\frac{1}{k^2-m_Z^2}\left(g^{\mu\nu}-\frac{k^{\mu}k^{\nu}}{k^2}\right)
\label{eqAmpZ}
\end{align}

Considering the implicit regularization, the Lorentz algebra must be performed prior the isolation of the
BDI's. If it is considered, alternatively, we can split the integrand of the second integral in eq. (\ref{eqAmpZ}) in a basic quadratic divergent integral with Lorentz indices and a logarithmic divergent one:
\begin{align}
&\int_k \frac{k^{\mu}k^{\nu}}{k^2(k^2-m^2_Z)}\nonumber\\
&= \int_k \frac{k^{\mu}k^{\nu}}{(k^2-m^2_Z)^2}-m^2_Z\int_k \frac{k^{\mu}k^{\nu}}{k^2(k^2-m^2_Z)^2}\nonumber\\
&=I^{\mu\nu}_{quad}(m^2_Z)-m^2_Z I^{\mu\nu}_{log}(m^2_Z)-\frac{b}{8}g^{\mu\nu}m^2_Z,
\label{eqZ}
\end{align}
where $b\equiv\frac{i}{(4\pi)^2}$.

Now, if eq (\ref{eqZ}) is inserted in eq. (\ref{eqAmpZ}) and the result is contracted with the metric, we find:
\begin{align}
&\frac{m_Z^2}{\nu^2}(I_{quad}(m_Z^2)+\upsilon_2)\nonumber\\
&+\frac{m_Z^2}{\nu^2}\left(\frac{m^2_Z}{2} I_{log}(m_Z^2)
-\frac{m^2_Z}{2}\upsilon_0 +\frac{b}{4}m^2_Z \right),
\end{align}
which means that the Lorentz algebra not performed as the first step not only generates additional surface terms but also a logarithmic divergent integral plus finite terms in the UV limit. Besides, it changes the coefficient of the quadratic divergence and as can be shown, this alters the coefficients of the particle masses in the Veltman's condition.

Based on the gauge symmetry condition, previously obtained in eq. (\ref{eqpi}), we can consider $\upsilon_2=0$. This changes the quadratic divergent behavior of the amplitude to $\frac{m_Z^2}{\nu^2}I_{quad}(m_Z^2)$ instead of $\frac{3m_Z^2}{2\nu^2}I_{quad}(m_Z^2)$ obtained by performing the Lorentz contraction prior the regularization. The same situation happens for the tadpole diagram with a $Z$ loop and for two corresponding 1-loop diagrams with a gauge boson $W^{\pm}$ loop. 

Nevertheless, even if a quadratic surface term is spuriously generated, it is also null as in the gauge condition if we apply the soft scale symmetry breaking regulators of section \ref{s4}. The choice of the explicit regularization presented in that section is not incompatible with gauge symmetry as we are going to show below. The surface terms are not arbitrary if we choose an explicit regularization. However, regardless of the choice of the regulator, the most general value for a quadratic surface term with two indices can be obtained using eqs. (\ref{dif1}) and (\ref{ts1}):
\begin{align}
\Upsilon_2^{\mu\nu}=g^{\mu\nu}I_{quad}(m^2)-2I^{\mu\nu}_{quad}(m^2)=g^{\mu\nu}\upsilon_2
\nonumber\\
\Rightarrow \upsilon_2=(c_2-c_2')\Lambda^2+(c_1-c_1')m^2,
\label{eqSTQ}
\end{align}
where we also had to use eq. (\ref{eqPARAM}) and the parametrization for a quadratic BDI with indices
\begin{align}
I^{\mu\nu}_{quad}(m^2)&=\frac{b}{2}g^{\mu\nu}c_2'\Lambda^2 \nonumber\\
&+\frac{b}{2}g^{\mu\nu}m^2\left(1+c_1'+\ln\frac{\Lambda^2}{m^2}\right) 
\label{param2}
\end{align}
that can be easily built with the use of the regularization independent differential
equations for BDI's with two indices:
\begin{align}
&\frac{\partial I_{quad}^{\mu\nu}(m^2)}{\partial m^2}=2I_{log}^{\mu\nu}(m^2),\nonumber\\
&\frac{\partial I_{log}^{\mu\nu}(m^2)}{\partial m^2}=\frac{-b}{4m^2}g^{\mu\nu}.
\label{eqDIFFEQ2}
\end{align}

When the general parametrization for $I^{\mu\nu}_{quad}(m^2)$ and $I_{quad}(m^2)$ are applied in the first line of eq. (\ref{eqSTQ}), with the use of eqs. (\ref{eqPARAM}) and (\ref{param2}), it leads to the second line of (\ref{eqSTQ}). This equation is another way to see that the arbitrary coefficients $c_2$
and $c_2'$ of the quadratic divergences are related to a surface term as presented in another form in eq. (\ref{eqSTC2}) with the use of an explicit regularization.

On the other hand, we can explicitly evaluate the quadratic integral
\begin{equation}
I^{\mu\nu}_{quad}(m^2)=\int_k \frac{k^{\mu}k^{\nu}}{(k^2-m^2)^2}.
\end{equation}

We apply the soft scale symmetry breaking regularization of section \ref{s4} and the
symmetric integration $k^{\mu}k^{\nu}\rightarrow \frac{1}{4}g^{\mu\nu}k^2$ in
order to get rid of the indices. This leads us to the following quadratic integral in the euclidean space:
\begin{eqnarray}
&I_{quad}^{\mu\nu}(m^2)=\frac{-i\Lambda^2}{64\pi^2}g^{\mu\nu}\int^{\infty}_0
dx \frac{x^2R(x)}{(x+a)^2}\nonumber\\
&=\frac{-i\Lambda^2 g^{\mu\nu}}{64\pi^2}\int^{\infty}_0 dx\left(
R(x)-\frac{2aR(x)}{x+a}+\frac{a^2R(x)}{(x+a)^2}\right),
\end{eqnarray}
where $x=\frac{k^2_E}{\Lambda^2}$ and $a=\frac{m^2}{\Lambda^2}$ as before.

We have the same integrals as before except for the last integral, whose result is
\begin{equation}
\int^{\infty}_0 dx \frac{R(x)}{(x+a)^2}=1+\frac{1}{a}-(2+a)e^a \Gamma(0,a),
\end{equation}
where $ \Gamma(0,a)=-\gamma_E-\ln a+a +...$ for $a<<1$.

Again, we see that we get $c_2'=0$ plus additional terms that vanish in the classical
limit $m\rightarrow 0$, which implies that $\upsilon_2=0$. So, if the regulator does not introduce a hard breaking of scale invariance, this is compatible with the value required by the gauge Ward identity for the quadratic surface term ($\upsilon_2=0$).

In the usual implicit regularization approach, the surface terms are not computed since no explicit regulator
is chosen. Then, such surface terms remain in the final result of the amplitudes, as in the case of the vacuum polarization tensor presented in eq. (\ref{eqpi}), and can be fixed by requiring a symmetry via a
Ward identity or phenomenology \cite{Jackiw2}. In this section, we explicitly computed $\upsilon_2$ because
the regulator was chosen. Nevertheless, we see that this computation makes soft breaking of scale symmetry compatible with the choice for the surface term imposed by the gauge Ward identity.

\section{Geometric Interpretation}
\label{s6}

We know that a sharp cutoff regularization corresponds to the volume of a hypersphere
in four dimensions in the evaluation of a divergent integral. That is why if this regulator is chosen, it hardly breaks scale invariance:
\begin{align}
&I_{quad}(m^2)\sim\int^{\infty}_0 dk^2_E R(k^2_E/\Lambda^2)\nonumber\\
&\rightarrow V_{\mathcal{N}}= \int^{\infty}_0 R(x) dx,
\end{align}
where $V_{\mathcal{N}}$ is the normalized radial hyper-sphere volume. We have $V_{\mathcal{N}}=1$ for sharp cutoff regularization or $R(x)=R_0(x)=L_0(x)e^{-x}$.

Unlike the Laguerre polynomial of degree zero, the regulator corresponding to $L_1(x)$
produces a vanishing normalized volume:
\begin{eqnarray}
 V_{\mathcal{N}}&= \int^{\infty}_0 R_1(x) dx=\int^{1}_0 R_1(x) dx+\int^{\infty}_1 R_1(x) dx\nonumber\\
&=V_{\mathcal{N}}^++V_{\mathcal{N}}^-=0,
\end{eqnarray}
where $V_{\mathcal{N}}^+$ and $V_{\mathcal{N}}^-$ are the positive and negative contributions for the hyper-volume, respectively. 

\begin{figure}[htb]
\includegraphics[scale=0.8]{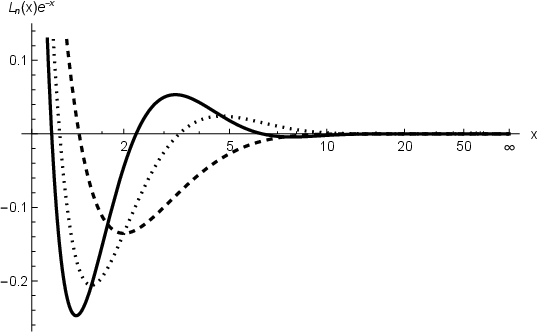}
\caption{Plot of the first three Laguerre polynomials damped by the exponential, skipping $n=0$. The
dashed, dotted and full lines stand for $L_1(x)$, $L_2(x)$ and $L_3(x)$, respectively.}
\label{fig3}
\end{figure}

It is easy to see why $V_{\mathcal{N}}$ is null if we analyze the behavior of the Laguerre polynomials 
suppressed by the exponential as presented in the graphic of figure \ref{fig3} for the first three Laguerre polynomials suppressed by the exponential $e^{-x}$, excluding $n=0$. We can check that there is a convergence of the regulator function $R_n(x)$ to zero for $x\sim10$, which means that $k_E$ and $\Lambda$ can differ up to one order of magnitude.

Of course, these normalized volumes cancel each other for the other Laguerre polynomials and the other corresponding regulators $R_n(x)$ for $n>0$. The implication of this is that
instead of only one hyper-solid with large radius, we have a combination of two
hyper-solids. For instance, considering the case for $n=1$ we have:
\begin{align}
\int^{\infty}_0 dk^2_E R(k^2_E/\Lambda^2)=&\int^{\infty}_0 dk^2_E e^{-k^2_E/\Lambda^2}\nonumber\\
&-\int^{\infty}_0 dk^2_E \frac{k^2_E}{\Lambda^2}e^{-k^2_E/\Lambda^2}.
\end{align}

It is important to notice that the regulator $R(x)$ does not produce a hyper-sphere with null hyper-volume. It is instead a difference between two hyper-regions. As we are going to show below, $R(x)$ is a continuous
sum of combined weighted cutoff's. Another way to interpret and find out the regulator function $R(x)$ is to
try to generalize the idea of a simple cutoff. We know that,
\begin{align} 
I_{quad}(m^2)\sim \Lambda^2=&\int^{\infty}_0 dk^2_E \theta \left(1-\frac{k^2_E}{\Lambda^2}\right)\nonumber\\
&=\Lambda^2 \int^{\infty}_0 dx\theta(1-x),
\end{align}
where $\theta(1-x)$ is the Heaviside step function.

So, for a sharp cutoff, we have $R(x)=\theta(1-x)$. We could instead have a 
continuous sum of cutoff's with different weights $w(y)$:
\begin{equation}
R(x)=\int^{\infty}_0  w(y)\theta(y-x)dy,
\label{eqSUP}
\end{equation}
where $\theta(y-x)=1$ if $y>x$ and $\theta(y-x)=0$ otherwise. We can recover the
well known sharp cutoff case by choosing $ w(y)=\delta(y-1)$.

The condition $y>x$ implies in $k_E<\Lambda \sqrt{y}$, which shows the variable maximum radius of each corresponding hypersphere, each of them contributing with weight $w(y)$.

If we split the integral in eq. (\ref{eqSUP}), we have
\begin{equation}
R(x)=\int^{\infty}_x  w(y) dy,
\label{eqSUP2}
\end{equation}
which implies that $R'(x)=-w(x)$ since $\int^{\infty}_x dR(y)=R(\infty)-R(x)=-\int^{\infty}_x  
w(y) dy$. 

Now, considering the first degree Laguerre polynomial, we have 
$R(x)=(1-x)e^{-x}$ and as a consequence $R'(x)=-(2-x)e^{-x}=-w(x)$ and finally
\begin{equation}
R(x)= \int^{\infty}_0 (2-y)e^{-y}\theta(y-x)dy.
\end{equation}

In this way, because $ \int^{\infty}_0 (2-y)e^{-y}\theta(y-x)dy=\int^{\infty}_x (2-y)e^{-y}dy$,
hyper-spheres weighted in region $0<y<2$ contribute to the hyper-volume with positive
sign while hyper-spheres in the region  $y>2$ contribute to the hyper-volume with negative sign in such a way that the hyper-regions cancel each other. There are much
more possibilities in the region corresponding to  $\infty>y>2$. However, this is compensated by the suppression caused by the exponential.

Analogously, we can build different weight functions for each one of the Laguerre polynomials. The general situation is $R'(x)=(L'_n(x)-L_n(x))e^{-x}=-w(x)$, that leads to
\begin{equation}
R(x)=\int^{\infty}_0 (L_n(y)-L'_n(y))e^{-y}\theta(y-x)dy,
\end{equation}
where we can again check that the hyper-region has always positive value for $n=0$.

\section{ 1-loop corrections to the Higgs propagator}
\label{s7}

The Standard Model of particle physics is widely tested \cite{PDG} and describes successfully all interactions but gravity. Nevertheless, there are still unanswered questions not explained by it, like the strong CP problem \cite{strongCP}, the existence of dark matter and dark energy \cite{DMDE}, the neutrino masses and neutrino oscillations \cite{Neutrinos} or the fine-tuning problem \cite{Susskind, Veltman}. As it is well-known, the tree-level correction to the Higgs boson mass can receive large corrections from the 1-loop diagrams if the limit of validity of the SM is assumed to occur at Planck scale. This would require an incredible fine-tunning so that the 1-loop correction of the Higgs mass is precisely canceled by the tree-level mass resulting in the Higgs mass measured in the laboratory. As an attempt to solve this, Veltman proposed a symmetry between masses that could cancel the quadratic correction \cite{Veltman}. This would avoid the fine-tunning but at the same time, the masses of particles should be related so that no fine-tuning is required. This relation is questioned to be valid for all energy scales \cite{Chaichian}. However, Veltman's proposal was the first step towards the construction of super-symmetric models. Veltman's condition can also be derived in a model independent way, such as the scale expected for new physics can be enlarged for $19\ TeV$\cite{CPC}.

At the same time, gauge symmetry is one of the main criteria to include field interactions in a quantum field theory. It is 
well-known that it is broken in the Standard Model by the Higgs mechanism. This process gives masses for the gauge bosons and assures
that the theory is renormalizable. Nonetheless, there still remains gauge symmetry since the photon remains massless. In other words,
a larger gauge symmetry is broken in a smaller one, {\it i.e.} $SU(2)\times U(1)\rightarrow U(1)$, after the vacuum expectation value of the scalar field assumes
a critical value. This gauge symmetry breaking is physical unlike the spurious symmetry breaking caused by regularization schemes. Although dimensional regularization \cite{DR, Bollini} is applied to prove the renormalizability of the SM, it is usually not considered to discuss the fine-tuning problem since quadratic divergences are absent with the use of this method. The sharp cut-off regularization, on the other hand, spuriously break gauge symmetry and it is usually considered to derive the 1-loop correction to the Higgs boson mass. The fact that gauge symmetry is broken by a regularization is not a problem at all as long as the spurious gauge symmetry breaking term is removed in the process of renormalization. This procedure is followed, for instance, in lattice regularization, where Lorentz symmetry is known to be broken spuriously and to be restored in the renormalization process. However, it is not always possible to identify exactly what is the spurious breaking term. Of course, it is also possible to build SuSy 
models to avoid all scale symmetry breaking quadratic divergences.

Now, returning to the previous discussions about soft scale symmetry breaking regulators, 
it is possible to sum all the quadratic divergent contributions in a single function using the scale relation in order to change the mass dependence of the basic divergent quadratic integral:

\begin{align}
I_{quad}(m^2_2) &= I_{quad}(m^2_1) + m^2_2 I_{log} (m^2_2)  
\nonumber\\
&- m^2_1 I_{log}(m^2_1)+ b(m^2_2 - m^2_1).
\label{srq}
\end{align}

Using eq. (\ref{srq}) and the result of the amplitudes of each quadratic divergent diagram we find that the full quadratic divergent contribution in the 1-loop amplitudes of the Higgs propagator is given by:
\begin{align}
&\frac{-6}{\nu^2}[m^2+m^2_Z+2m^2_W-4m^2_t]I_{quad}(m^2)+...,
\label{eqRES}
\end{align}
where $m$, $m_W$, $m_Z$ and $m_t$ are the Higgs, $W^{\pm}$,  $Z$ and top quark 
masses respectively. the ellipsis stands for logarithmic divergences and other finite 
contributions in the UV limit.

The OMS renormalization together with the full contribution presented in eq. (\ref{eqRES})
lead us to the correction to the Higgs mass:
\begin{align}
&m^2_r=m^2-\frac{3}{8\pi^2\nu^2}(m^2+m_Z^2+2m_W^2-4m^2_t)
\nonumber\\
&\times\left(c_2\Lambda^2+(1+c_1)m^2+m^2\ln\frac{\Lambda^2}{m^2}\right)+...,
\label{eqHM}
\end{align}
where we can check the Veltman condition by making the first parenthesis null \cite{Veltman}. We can
try to restore the classical limit in this equation by taking $m\rightarrow 0$. If we do so,
the only term that breaks scale symmetry using eq. (\ref{eq18}) is

\begin{align}
\Theta^{\mu}_{\ \ \mu}&=-\frac{3}{8\pi^2\nu^2}(m_Z^2+2m_W^2-4m^2_t)c_2\Lambda^2\nonumber\\
&\times\Phi(x)^{\dagger}\Phi(x).
\label{eqEMM}
\end{align}

Since the coefficients of eq. (\ref{eqHM}) depend on the soft scale symmetry breaking regularization chosen in sec. \ref{s4}, it is possible to recover scale symmetry in the classical limit. We checked that if the regulator depends on the Laguerre polynomial of degree 1, then $c_2=0$ and $c_1=-2-\gamma_E$, which means
that $\Theta^{\mu}_{\ \ \mu}=0$ for $m\rightarrow 0$. Furthermore, $c_2=0$ for all Laguerre polynomials of degree $n\neq0$, but the numerical value of $c_1$ can change for these other regulators, although the logarithmic dependence on $\Lambda$ always cancel for $n>1$ in $c_1$ as well. The remaining corrections to the Higgs tree-level mass that have logarithmic dependence do not introduce a hard breaking of scale symmetry because they are multiplied by $m^2$ as already observed in \cite{IJTP, AOP}. 

Eq. (\ref{eqEMM}) and the absence of the term $c_2\Lambda^2$ should be interpreted as a consequence of the
use of a soft scale symmetry breaking regulator. This means that there is no quadratic correction to the 
Higgs mass caused by the regulator but not the absence of a quadratic sensitivity in a general UV completion
because we can not claim that the SM is arbitrarily valid (up to Planck Scale for instance) and 
that there is no other robust theories to be considered, including the ones with heavy fields that couple to the Higgs.

Furthermore, it is worth highlighting that eq. (\ref{eqEMM}) is a result obtained at 1-loop for this class
of regulators. The 2-loop issues are going to be discussed in the next section. However, the full mass correction to the Higgs mass at 2-loop is highly non-trivial and although there are some patterns corresponding to common topologies, there are hundreds of diagrams to regularize. 

\section{Application beyond 1-loop} 
\label{s8}

\begin{figure}[htb]
\centering
\includegraphics[trim= 0mm 0mm 0mm 5mm, scale=0.75]{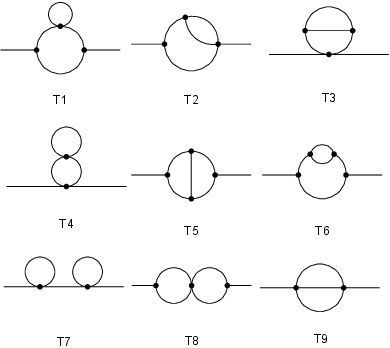}
\caption{2-loop topologies to the Higgs propagator.}
\label{fig6}
\end{figure}

In this section, we discuss the application of the Laguerre explicit regulators in amplitudes beyond
1-loop. Some topologies lead to divergent structures that are products of the same BDI's that appear at
1-loop order and can be easily regulated as before. This is the case of topologies T1, T4, T6, T7 and T8
presented in figure \ref{fig6}. The more interesting cases are the overlapping divergences, like the ones
that appear in topologies T2, T3, T5 and T9. These topologies lead to different BDI's. The ones that appear at 2-loop are \cite{2loopQCD}:
\begin{align}
& I_{log}^{(2)}(m^2)=\int_k\frac{1}{(k^2-m^2)^2}\ln \left( -\frac{k^2-m^2}{\mu^2}\right), \nonumber\\
& I_{quad}^{(2)}(m^2)=\int_k\frac{1}{(k^2-m^2)}\ln \left( -\frac{k^2-m^2}{\mu^2}\right),
\label{2LBDIS}
\end{align}
where $\mu^2$ is a mass scale. The form of the possible divergences is summarized in table \ref{tab01} for each kind of topology presented in figure \ref{fig6}.

\begin{table}
\centering
 \caption{Expected BDI's for each topology at 2-loop presented in figure \ref{fig6}.}
 \begin{tabular}{cc}
  \hline
  Topology & Possible divergent structures \\
  \hline
  T1 & $I_{quad}(m^2)$ \\
  \hline
  T2 & $I_{log}(m^2), I^2_{log}(m^2), I^{(2)}_{log}(m^2)$ \\
  \hline
  T3 &  $I^2_{log}(m^2), I^{(2)}_{log}(m^2)$ \\
  \hline
  T4 &  $I_{log}(m^2)I_{quad}(m^2)$ \\
  \hline
  T5 & 0\\
  \hline
  T6 & $I_{log}(m^2)$ \\
  \hline
  T7 & $I^2_{quad}(m^2)$\\
  \hline
  T8 & $I_{log}(m^2), I^2_{log}(m^2)$\\
  \hline
  T9 & $ I_{log}(m^2), I_{log}(m^2)I_{quad}(m^2), I^{(2)}_{quad}(m^2)$ \\
  \hline 
 \end{tabular}
\label{tab01} 
\end{table}

The logarithmic BDI that appears at 2-loop as a function of $x=k_E^2/\Lambda^2$ can be written as
\begin{align}
 &I_{log}^{(2)}(m^2)=b\int^{\infty}_0 dx \frac{x}{(x+a)^2}\ln \left((x+a)\frac{\Lambda^2}{\mu^2}\right)
 \nonumber\\
 &=b\left(\int^{\infty}_0 dx \frac{x \ln (x+a)}{(x+a)^2}+\int^{\infty}_0 dx\frac{x \ln (\Lambda^2/\mu^2)}{(x+a)^2}\right),
\label{BDI2L}
\end{align}
with $a=m^2/\Lambda^2$ as previously presented.

The second integral in eq. (\ref{BDI2L}) is the same that appeared in eq. (\ref{BDI1L}) and can be regulated
by any of the functions $R_n(x)=L_n(x)e^{-x}$, as before. While for the first integral, we can consider the
expansion $\frac{x}{(x+a)^2}\ln (x+a)\approx \frac{\ln x}{x}+\mathcal{O}(a)$ for $a<<1$. Now, from
$\int^{\infty}_0 dx \frac{\ln x}{x}$, we see that there are two sources of divergences in the integral, one
coming from $1/x$ and another from $\ln x$. None of the Laguerre polynomials can get rid of these divergences
alone. However, we know that they form an orthogonal basis that can be used to write another regulator function $R_n(x)=P_n(x)e^{-x}$. The linear combination that can remove these divergences is 
$P_1(x)=L_0(x)-L_1(x)$, so that we have:
\begin{equation}
\int^{\infty}_0 \frac{dx}{x} \ln x \rightarrow \int^{\infty}_0 \frac{dx}{x} (L_0-L_1)e^{-x}\ln x=-\gamma_E.
\end{equation}   

This combination of Laguerre polynomials however does not keep the correct IR regime because now $P_n(0)=0$. In this way, the Laguerre polynomials are good in not introducing additional breaking of scale symmetry and in the removal of some IR divergences. Nevertheless, some physical content of the starting amplitude might be altered in the IR regime because $P_n(0)=1$ is not valid anymore. It is not possible at the same time to remove the divergence and not to alter the IR regime for the logarithmic BDI 
$I^{(2)}_{log}(m^2)$. In any case, these IR structures that appear at 2-loop do not break scale symmetry in a hard way because they are dimensionless.

Unlike the 2-loop logarithmic BDI, the quadratic one can break scale symmetry in a hard way depending on
how it is regulated. Considering $I^{(2)}_{quad}(m^2)$ in eqs. (\ref{2LBDIS}) written in terms of $x$, we have
\begin{align}
&I^{(2)}_{quad}(m^2)=-b\Lambda^2\int^{\infty}_0\Big( \ln \frac{\Lambda^2}{\mu^2}+ \ln (x+a)\nonumber\\ 
&-\frac{a}{x+a}\ln \frac{\Lambda^2}{\mu^2} -\frac{a}{x+a}\ln(x+a) \Big) dx\nonumber\\ 
&\rightarrow -b\Lambda^2\Big( \ln \frac{\Lambda^2}{\mu^2}\int^{\infty}_0 dx R(x)\nonumber\\
&+\int^{\infty}_0 dx R(x) \ln (x+a)-a\int^{\infty}_0\frac{dx R(x)}{x+a}\ln \frac{\Lambda^2}{\mu^2}\nonumber\\
&-a\int^{\infty}_0\frac{dx R(x)}{x+a}\ln(x+a) \Big),
\end{align}

The first integral is the same appearing at 1-loop and the second one can not be regulated by only one
Laguerre regulator. Again, we try a linear combination $P_2(x)=A_0L_0(x)+A_1L_1(x)+A_2L_2(x)$ such that
$P_2(0)=1$ and $\int^{\infty}_0 dx R_2(x) \ln x=0$. These requirements lead us to $A_0=0$, $A_1=-1$ and $A_2=2$. Thus, the 2-loop BDI $I^{(2)}_{quad}(m^2)$ requires the regulator function $P_2(x)=(-L_1(x)+2L_2(x))e^{-x}=(x^2-3x+1)e^{-x}$. The last two integrals does not require a combination 
of Laguerre polynomials and we can take $R_0(x)=e^{-x}$ for simplicity. Considering the last integral, notice that this time there is no issue related to $\int^{\infty}_0 dx\frac{\ln x}{x}$ in the massless limit because it is multiplied by $a$. The full result of the regulated integral is therefore:
\begin{align}
I_{quad}^{(2)}(m^2)&=b m^2\left(1-\ln\frac{\Lambda^2}{\mu^2}\right)\left( \gamma_E+\ln \frac{m^2}{\Lambda^2}\right)\nonumber\\
&+\frac{b}{2}m^2\left(\gamma^2_E+\frac{\pi^2}{6}-\ln^2\frac{m^2}{\Lambda^2}+2\right),
\label{eqIQ2}
\end{align}
where we can check that all terms vanish in the massless limit.

\begin{figure}[htb]
\centering
\includegraphics[trim= 0mm 0mm 0mm 5mm, scale=0.7]{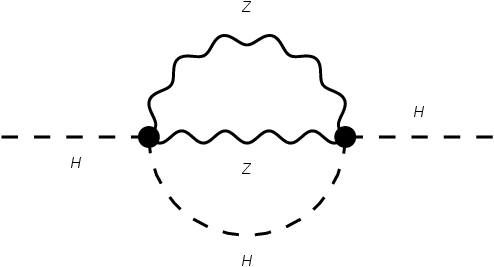}
\caption{Overlapped 2-loop correction to the Higgs propagator.}
\label{fig7}
\end{figure}

As an application, let us consider the 2-loop correction to the Higgs propagator coming from the 
overlapping diagram of figure \ref{fig7}. After performing Lorentz algebra the amplitude can be written
as
\begin{align}
&\frac{9i\nu^2 g^4}{4 c^4_W}\int_l\int_k \frac{1}{(k^2-m^2_Z)(l^2-m^2_Z)[(k-q)^2-m^2]}\nonumber\\
&=\frac{36i m^4_Z}{\nu^2}\int_l \frac{1}{(l^2-m^2_Z)}(I_{log}(m^2_Z)\nonumber\\
&-b(m^2-m^2_Z)\iota^{(1)}_0-b Z^{(1)}_0(q^2,m_Z^2)),
\label{eq2LH}
\end{align}
where again $m$ and $m_Z$ are the Higgs and the $Z$-boson masses, respectively, $c_W=\cos \theta_W$, 
$q=l+p$, $\iota^{(1)}_0=\int dx\frac{(1-x)}{\Delta^{(1)}}$, $Z^{(1)}_0(q^2,m_Z^2)=\int^1_0 dx \ln \left[\frac{\Delta^{(1)}}{m^2_Z}\right]$ and $\Delta^{(1)}=m^2x+m^2_Z(1-x)-q^2 x(1-x)$.

Because of the interaction, the amplitude is not proportional to the Higgs mass and when it is regulated it can spuriously break scale symmetry. In order to take the Higgs massless limit, we need to use the scale relation for a logarithmic BDI:
\begin{equation}
I_{log}(m^2_Z)=I_{log}(\mu^2)+b\ln\frac{\mu^2}{m_Z^2},
\end{equation}
which turns eq. (\ref{eq2LH}) in the following result in the massless limit

\begin{align}
&\frac{36i m^4_Z}{\nu^2} \Big( I_{quad}(m^2_Z)(I_{log}(\mu^2 )+2b)\nonumber\\
&+b\int_l \frac{1}{(l^2-m^2_Z)}\ln \left[\frac{m_Z^2-(p+l)^2}{\mu^2}\right]\Big).
\label{eq2LH2}
\end{align}

The first term of the eq. (\ref{eq2LH2}) has the usual 1-loop BDI's and it can be regulated as presented in section \ref{s4}. The second one contains the quadratic BDI of eq. (\ref{2LBDIS}). In order to see this we
can expand the logarithm in the UV limit:
\begin{align}
&\ln \left[\frac{m_Z^2-(p+l)^2}{\mu^2}\right]=\ln \frac{-l^2+m^2_Z}{\mu^2}+\frac{2l\cdot p}{l^2-m^2_Z}\nonumber\\
&+\frac{p^2}{l^2-m^2_Z}-\frac{(2l\cdot p)^2}{2(l^2-m^2_Z)^2}-\frac{p^4}{2(l^2-m^2_Z)^2}+...
\end{align}

The first term of the expansion leads to $I^{(2)}_{quad}(m^2_Z)$ and according to eq. (\ref{eqIQ2}) implies in a soft breaking of scale symmetry. The second term is null because it leads to an odd integral in $l$. The third term is null in the massless limit with the on-mass-shell condition $p^2=m^2$. The second order term leads to $m^2I_{log}(m^2_Z)$ and higher order terms are finite, being proportional to $m^2(m^{2n}/(\Gamma(n)m^{2n}_Z)$ and they can be neglected when the order $n$ increases.

The first term of the expansion leads to $I^{(2)}_{quad}(m^2_Z)$ and according to eq. (\ref{eqIQ2}) implies in a soft breaking of scale symmetry. The second term is null because it leads to an odd integral in $l$. The third term is null in the massless limit with the on-mass-shell condition $p^2=m^2$. The second order term leads to $m^2I_{log}(m^2_Z)$ and higher order terms are finite, being proportional to $m^2/\Gamma(n)$ and they can be neglected when the order $n$ increases.

In general, for new divergent structures of BDI's, we may need to consider another linear combination of the Laguerre polynomials in order to find the polynomial $P_n(x)$ in the regulator $R_n(x)=P_n(x)e^{-x}$. The
expansion of a $P_N(x)$ polynomial in a Laguerre basis is given by:
\begin{equation}
P_N(x)=\sum^N_{n=0} A_n L_n(x),
\label{eqBASIS}
\end{equation} 
where $A_n$ are arbitrary coefficients fixed by the requirements $P_N(0)=1$, $\Lambda^2\int^{\infty}_0 P_l(x)e^{-x} h_l(x) =0$ and $h_l(x)=\ln^{l-1}x$, being $l$ the number of loops.

The BDI's beyond 2-loop are also defined in implicit regularization. They are found after the use of the recursive relation in overlapped multi-loop amplitudes \cite{2loopQCD}:
\begin{align}
&I^{(l)}_{log}(\mu^2 )=\int_{k_l}\frac{1}{(k^2_l-m^2 )^2}\ln^{l-1}\left( -\frac{k^2_l-m^2}{\mu^2}\right)
\nonumber\\
&\text{and}\nonumber\\
&I^{(l)}_{quad}(\mu^2)=\int_{k_l}\frac{1}{(k^2_l-m^2 )}\ln^{l-1}\left( -\frac{k^2_l-m^2}{\mu^2}\right),
\end{align}
where $l$ is the number of loops.

These BDI's require the use of the Laguerre basis of eq. (\ref{eqBASIS}) with coefficients determined by the absence of quadratic cutoff and the condition $P_N(0)=1$.

\section{Conclusions and perspectives} 
\label{s9}

In this work, we show that it is possible to find out a family of regulators that softly break scale 
symmetry. These regulators are particularly relevant to situations where we are interested in finding out
the correct physical breaking of scale symmetry, {\it i. e.} the ones that can restore the classical 
limit when taking $m\rightarrow 0$ and $\beta\rightarrow 0$. This might have important
implications for the Higgs mass computation, although there are several alternative proposals
to remove the quadratic cutoff sensitivity associated with the regularization scheme, being the SuSy models the most popular ones.

It is interesting to notice that the soft scale symmetry breaking regulators depend on the Laguerre 
polynomials. However, the family of functions only needs to have the property $R_n(0)=1$, with $P_n(x)$ 
being orthogonal. Of course the Laguerre polynomials are not the only ones with this property. Also, this 
approach is dependent on the original version of implicit regularization \cite{Orimar0} if we want to avoid 
the appearance of additional surface terms. If the amplitude is split in BDI's plus finite pieces, we are 
allowed to perform shifts in the finite integrals or logarithmic divergences via Feynman parametrization. On 
the other hand, if the starting amplitude has superficial divergent degree higher than logarithmic, 
additional surface terms can be generated when the shifts in the integrated momenta are performed during the 
Feynman parametrization.

The elimination of the term $c_2\Lambda^2$ via the Laguerre regularization demonstrates the absence of the
Higgs mass sensitivity associated to the regularization scheme. However, if our purpose is to investigate
if this Laguerre regularization supplied by the scale symmetry argument also protects the Higgs against heavy physical scales($M$ for instance), we need an extension of the SM (SuSy versions or not) with degrees of freedom so that $M>>m$ \cite{Hamada}.

The soft scale symmetry breaking Laguerre regularization also make us remind of the Hydrogen atom, where a 
similar situation occurs. The quantization procedure applied to the truncation of a series leads us to solutions to the radial equation that are the associated Laguerre polynomials. However, these polynomials are part of the wave function while in this work they are used as regulators in order to make the amplitude finite or free of quadratic divergences. Nevertheless, the Laguerre polynomials show up in radial physical problems, where in both situations (Hydrogen atom and soft scale symmetry breaking) we need orthogonality of the polynomials in the line $[0,\infty[$.

Furthermore, as a prospect, it would be interesting to investigate if the explicit symmetry breaking terms 
of SuSy models, in particular the non-soft SuSy breaking ones \cite{Grisaru2}, lead to quantum corrections that can be regulated in this approach, since these terms do not necessarily preserve scale symmetry in the classical limit, including the soft SuSy breaking ones.

\backmatter

\bmhead{Supplementary information}

\bmhead{Acknowledgements}

We thank the referee for the careful reading of our manuscript and for the constructive comments and suggestions.


\section*{Declarations}

\begin{itemize}
\item Data availability: This manuscript has no associated data or the data will not be deposited.
[Author's comment: Data sharing not applicable to this article as no datasets were generated or analysed
during the current study]. 
\item Code availability: Code/software will be made available on reasonable request. [Author's comment: The code generated or analysed in the preparation of the manuscript is available from the author on reasonable request.]  
\end{itemize}

\noindent

\bigskip

\begin{appendices}

\section{Result of the integrals} 
\label{A2}

The results of all finite and regularized divergent integrals for $\Pi^{\mu\nu}$ and
some of the 1-loop corrections to the Higgs propagator are listed below:

\be
\int_k \frac{k^{\mu}k^{\nu}}{k^2(k^2-m^2)^2}=\frac{1}{4}g^{\mu\nu}(I_{log}(m^2)-\upsilon_0)
+\frac{b}{8}m^2g^{\mu\nu},
\ee
\\
\be
\int_k \frac{1}{[(k+p)^2-m^2]}=I_{quad}(m^2)-p^2\upsilon_0, 
\ee
\be
\int_k \frac{1}{(k^2-m^2)[(k+p)^2-m^2]}=I_{log}(m^2)-b Z_0, 
\ee
\\
\begin{align}
\centering
&\int_k \frac{k^{\mu}}{(k^2-m^2)[(k+p)^2-m^2]}\nonumber\\
&=-\frac{1}{2}p^{\mu}(I_{log}(m^2)-\upsilon_0 -b Z_0),
\end{align}
\begin{align}
\centering
&\int_k \frac{k^{\mu}k^{\nu}}{(k^2-m^2)[(k+p)^2-m^2]}=\frac{1}{2}g^{\mu\nu}(I_{quad}(m^2)\nonumber\\
&-\upsilon_2 ) +\left(\frac{1}{3}p^{\mu}p^{\nu}-\frac{1}{12}p^{2}g^{\mu \nu} \right)( I_{log}(m^2) -bZ_0) \nonumber\\
&-\frac{1}{6}\xi_0(p^2g^{\mu\nu}+2p^{\mu}p^{\nu}) +\frac{1}{4}g^{\mu\nu}p^2 \upsilon_0 \nonumber\\
&-\frac{1}{3}b(p^2g^{\mu\nu}-p^{\mu}p^{\nu})\left(\frac{m^2}{p^2}Z_0 +\frac{1}{6} \right), 
\end{align}
where we used the following notations: $Z_n=Z_n(p^2,m^2)=\int_0^1 dx x^n \ln \frac{\Delta}{m^2}$, $b\equiv \frac{i}{(4\pi)^2}$ and $\Delta=\Delta(x)=m^2-p^2 x(1-x)$.
\end{appendices}

The following relation involving integrals in the Feynman parameters can be used to reduce finite
integrals:
\be
Z_k=\frac{k}{k+1} Z_{k-1} -\frac{k-1}{k+1}\frac{m^2}{p^2}Z_{k-2}-\frac{k-1}{k(k+1)^2},
\ee
where $k>0$ and $\iota_n=\int_0^1 dx \frac{x^n(1-x)}{\Delta(x)}$.

\end{document}